\documentclass[preprint,authoryear]{elsarticle}

\usepackage{amsmath,amssymb,mathabx}
\usepackage{stackengine,graphicx}
\usepackage{natbib}
\usepackage[dvipsnames]{xcolor}
\usepackage{bbm}
\usepackage{booktabs}

\usepackage{rotating}

\usepackage{algorithm}
\usepackage{algpseudocode}

\usepackage{collectbox}
\usepackage{bbm}
\usepackage{capt-of}
\usepackage{longtable}

\usepackage[center,font=small]{caption}

\usepackage{hyperref}
\hypersetup{
    colorlinks=true,
    linkcolor=blue,
    filecolor=magenta,      
    urlcolor=cyan,
    citecolor=black
}

\def\l{\left}
\def\r{\right}

\usepackage{array}
\newcolumntype{L}[1]{>{\raggedright\let\newline\\\arraybackslash\hspace{0pt}}m{#1}}
\newcolumntype{C}[1]{>{\centering\let\newline\\\arraybackslash\hspace{0pt}}m{#1}}
\newcolumntype{R}[1]{>{\raggedleft\let\newline\\\arraybackslash\hspace{0pt}}m{#1}}

\date{\today} 

\begin{document}

   \begin{frontmatter}

     \title{Aggregating predictions from experts: \\ a scoping review of statistical methods, experiments, and applications}

     \author[1]{Thomas~McAndrew \corref{cor1}}
     \ead{mcandrew@umass.edu}
     
     \author[1]{Nutcha~Wattanachit}
     \ead{nwattanachit@umass.edu}

     \author[1]{G.~Casey~Gibson}
     \ead{gcgibson@umass.edu}
     
     \author[1]{Nicholas~G.~Reich}
     \ead{nick@schoolph.umass.edu}
     
     \cortext[cor1]{Corresponding author}
     \tnotetext[funding]{This work has been supported by the National Institutes of General Medical Sciences (R35GM119582), the Defense Advanced Research Projects Agency, and the Centers for Disease Control and Prevention (1U01IP001122).
The content is solely the responsibility of the authors and does not necessarily represent the official views of NIGMS, the National Institutes of Health, the Defense Advanced Research Projects Agency, or and the Centers for Disease Control and Prevention.}

      \address[1]{Department of Biostatistics and Epidemiology, School of Public Health and Health Sciences, University of Massachusetts at Amherst, Amherst, Massachusetts, USA}

\begin{abstract}
  Forecasts support decision making in a variety of applications.
  Statistical models can produce accurate forecasts given abundant training data, but when data is sparse, rapidly changing, or unavailable, statistical models may not be able to make accurate predictions.
  Expert judgmental forecasts--—models that combine expert-generated predictions into a single forecast—--can make predictions when training data is limited by relying on expert intuition to take the place of concrete training data.
  Researchers have proposed a wide array of algorithms to combine expert predictions into a single forecast, but there is no consensus on an optimal aggregation model.
  This scoping review surveyed recent literature on aggregating expert-elicited predictions.
  We gathered common terminology, aggregation methods, and forecasting performance metrics, and offer guidance to strengthen future work that is growing at an accelerated pace.
\end{abstract}


\end{frontmatter}

\section{Introduction}

Forecasting presents decision makers with actionable information that they can use to prevent (or prepare for) economic~\citep{shin2013robust,huang2016improving,mak1996aggregating}, engineering~\cite{guangliang1996multi,zio1996use,neves2008life}, ecological~\cite{borsuk2004predictive,failing2004using,morales2017characterization,johnson2018making}, social~\cite{cabello2012combination,klas2010support,craig2001bayesian}, and public health burdens~\cite{evans1994use,alho1992estimating}.

Advances in computing power made statistical forecasts, models that take as input a structured data set and output a point estimate or probability distribution, a powerful tool~\citep{wang2016statistical,kune2016anatomy,al2015efficient}. 
Statistical models exploit correlations between data to find patterns, but when data is rapidly changing, sparse, or missing completely, the accuracy of these models can suffer. 
Judgmental forecasts attempt to overcome data limitations present in statistical models by eliciting predictions from experts~\citep{clemen1986combining,clemen1989combining,genest1986combining}.
Experts are able to make predictions without structured data, and instead, rely on their experience and contextual knowledge of the prediction task.
Expert forecasts are most readily found in finance, business, and marketing~\citep{seifert2013relative,shin2013robust,franses2011averaging,petrovic2006fuzzy,alvarado2017expertise,song2013combining,baecke2017investigating,baecke2017investigating,petrovic2006fuzzy,franses2011averaging,song2013combining,alvarado2017expertise,seifert2013relative,kabak2008aggregating}.
These fields focus on decision makers and their ability to make predictions from data that cannot easily be collected and fed to a statistical model.
Other areas of active research in expert opinion are quality assurance~\citep{klas2010support}, politics~\cite{hanea2018value,graefe2014accuracy,graefe2015accuracy,graefe2018predicting,cai2016simple,wang2018bayesian,satopaa2014probability,graefe2014combining}, economics~\cite{shin2013robust,huang2016improving,mak1996aggregating}, engineering~\cite{craig2001bayesian,tartakovsky2007probabilistic,neves2008life,ZIO1996127,BRITO201655,jin2007research,wang2008probabilistic,brito2012behavioral,hathout2016uncertainty,ren2002optimal}, sports~\cite{gu2016expert}, sociology~\cite{cabello2012combination,adams2009acceptability}, meteorological~\cite{abramson1996hailfinder}, ecological~\cite{johnson2018making,borsuk2004predictive,failing2004using,cooke2014out}, environmental science~\cite{morales2017characterization,mantyka2014understanding,li2012preliminary,zio1997accounting}, and public health~\cite{alho1992estimating,EVANS199415,jana2019interval,kurowicka2010probabilistic}.
The diversity and breadth of applications underscore the importance of expert opinion in a wide variety of disciplines.

Research combining expert opinion to produce an aggregate forecast has grown rapidly, and a diverse group of disciplines apply combination forecasting techniques.
Cross-communication between different applied areas of combination forecasting is minimal, and as a result, different scientific fields are working in parallel rather than together.
The same mathematical ideas in combination forecasting are given different labels depending on application.
For example, the literature refers to taking an equally-weighted average of expert forecasts as: equal-weighting, unweighted, and 50-50 weighting.

This scoping review focuses on methods for aggregating expert judgments.
The aim is to survey the current state of expert combination forecasting literature, propose a single set of labels to frequently used mathematical details, look critically at how to improve expert combination forecasting research, and suggest future directions for the field. 

We map key terminology used in combining expert judgemental forecasts and consolidate related definitions.
A textual analysis of scoped articles highlights how combination forecasting techniques have evolved.
A prespecified list of questions was asked of every in-scope manuscript: whether point predictions or predictive densities were elicited from experts, methods of aggregating expert predictions, experimental design for evaluating combination forecasts and how forecasts were scored (evaluated).
We tabulated techniques for evaluating forecasts and condensed terms referring to the same evaluative metric.

Section~\ref{background} gives a brief historical background of combination forecasting and current challenges.
Section~\ref{sec.methods} describes our literature search, how articles were defined as in-scope, and our analysis.
Section~\ref{sec.results} reports results and section~\ref{sec.discussion} discusses common themes, terminology, advocates for key areas that need improvement, and recommends future directions for aggregating expert predictions.

\section{Background}
\label{background}

\subsection{Human judgmental forecasting}

Judgmental forecasting models---predictions elicited from experts or non-expert crowds and combined into a single aggregate forecast---have a long history of making well calibrated and accurate predictions~\citep{edmundson1990decomposition,bunn1991interaction,lawrence1992exploring,o1993judgemental}.
Advances in judgmental forecasting take two paths: building sophisticated schemes for combining predictions~\citep{clemen1989combining,clemen1999combining,clemen2008comment} and eliciting better quality predictions~\cite{ayyub2001elicitation,helmer1967analysis}. 

Initial combination schemes showed an equally-weighted average of human-generated point predictions can accurately forecast events of interest~\citep{galton1907vox}.
More advanced methods take into account covariate information about the forecasting problem and about the forecasters themselves (for example weighting experts on their past performance).
Compared to an equally-weighted model, advanced methods show marginal improvements in forecasting performance~\citep{fischer1999combining,mclaughlin1973forecasters,armstrong1985crystal,winkler1971probabilistic,clemen1989combining}.

In this work we will study combinations of expert predictions.
Combining non-expert predictions often falls into the domain of crowdsourcing, and crowdsourcing methods tend to focus on building a system for collecting human-generated input rather than on the aggregation method.

Past literature suggests experts make more accurate forecasts than novices~\citep{armstrong2001combining,armstrong1983relative,lawrence2006judgmental,spence1997moderating,alexander1995refining,french2011aggregating,clemen1999combining}.
Several reasons could contribute to this increased accuracy: domain knowledge, the ability to react to and adjust for changes in data, and the potential to make context-specific predictions in the absence of data~\citep{armstrong1983relative,lawrence2006judgmental,spence1997moderating,alexander1995refining}.
The increased accuracy of expert opinion led some researchers to exclusively study expert forecasts~\citep{armstrong2001combining,french2011aggregating,genre2013combining},
however crowdsourcing---asking large volumes of novices to make predictions and using a simple aggregation scheme---rivals expert-generated combination forecasts~\citep{howe2006rise,lintott2008galaxy,prill2011crowdsourcing}.
Whether or not expert or non-expert predictions are solicited, judgmental forecasting agrees that human judgment can play an important role in forecasting.

Judgmental forecasts can have advantages over statistical forecasting models. 
Human intuition can overcome sparse or incomplete data issues.
Given a forecasting task with little available data, people can draw on similar experiences and unstructured data to make predictions, whereas statistical models need direct examples and structured data to make predictions.
When data is plentiful and structured, statistical models typically outperform human intuition~\citep{meehl1954clinical,kleinmuntz1990we,yaniv1993judgmental}.
But whether a statistical or judgemental forecast is best depends on the circumstances. 

An understanding of the type of forecasts that models can produce and a mathematical description of a combination forecast can clarify how judgmental data, number of forecasters, and the combination scheme interact.

\subsection{A framework for combination forecasting}

Forecasting models can be statistical, mechanistic, or judgmental. 
We define a forecasting model $\mathcal{M}$ as a set of probability distributions over all possible events.
Each probability distribution is typically assigned a vector~$(\theta)$, called the model's parameters, that is used to differentiate one probability distribution from another
$\mathcal{M} = \l\{ P_{\theta} | \theta \in \Theta \r\}$,
where $P_{\theta}$ is probability distribution for a specific choice of $\theta$, and $\Theta$ are all possible choices of model parameters.

Models can produce two types of forecasts: point predictions or predictive densities. 
Point forecasts produce a single estimate of a future value~\citep{bates1969combination,granger1984improved} and are frequently used because they are easier to elicit from experts and early work was dedicated to combining specifically point forecasts~\cite{granger1984improved,bates1969combination,galton1907vox}.
Probabilistic forecasts are more detailed. 
They provide the decision maker an estimate of uncertainty (probability distribution) over all possible future scenarios~\citep{clemen1999combining,stone1961opinion,winkler1981combining,genest1986combining,winkler1968consensus,dawid1995coherent,ranjan2010combining,gneiting2013combining,hora2015calibration}. 
Probabilistic densities can be thought of as more general than point forecasts.
A point forecast can be derived from probabilistic forecast by taking, for example, the mean, median, or maximum a posteriori value.
A probabilistic density assigning all probability mass to a single value can be considered a point forecast.

A combination forecast aggregates predictions, either point or probabilistic, from a set of models and produces a single aggregate forecast~\citep{clemen1999combining,winkler1981combining,genest1986combining}.
Given a set of models $\mathcal{M}_{1},\mathcal{M}_{2},\cdots,\mathcal{M}_{N}$, a combination model $\mathcal{G}:\mathcal{M}_{1} \times \mathcal{M}_{2} \times \cdots \times \mathcal{M}_{N} \to \mathcal{F}$ maps the cartesian product of all models onto a single class of suitable probability distributions~\citep{gneiting2013combining}.
The goal of combination forecasting is to find an optimal aggregation function $G \in \mathcal{G}$.
Typically the model $\mathcal{G}$ is parameterized $\mathcal{G} = \l\{ G_{\upsilon} | \upsilon \in \Upsilon \r\} $ such that finding an optimal $G$ amounts to finding the parameter vector $\upsilon$ that produces an optimal forecast.

There are several ways to improve a combination model's forecasting ability.
Combination models can improve forecast accuracy by considering a more flexible class of aggregation functions~$(\mathcal{G})$.
Soliciting expert opinion (versus novices) can be thought of as improving individual forecasts $\mathcal{M}$ used as input into the combination model.
Crowdsourcing takes a different approach to improve forecast accuracy~\citep{howe2006rise,brabham2013crowdsourcing,abernethy2011collaborative,forlines2014crowdsourcing,moran2016epidemic}. 
These methods consider a simple class of aggregation functions $\mathcal{G}$ and collect a large number of human-generated forecasts $\mathcal{M}$.
By accumulating a large set of human-generated predictions, a crowdsourcing approach can create flexible models with a simple aggregation function.

This framework makes clear the goals of any combination forecasting model.
Some focus on improving individual models $\mathcal{M}$, others focus on more flexible aggregation functions~($\mathcal{G}$).
In this work we will consider combination forecasting models that include expert-elicited forecasts as their raw material and pursued building more flexible aggregations models.

\subsection{A brief timeline of existing work}

Francis Galton was one of the first to formally introduce the idea of combination forecasting. 
In the early 20th century, he showed aggregating point estimates from a crowd via an unweighted average was more accurate compared to individual crowd estimates~\citep{galton1907vox}.
Galton's work was empirical, but laid the foundation for exploring how a group of individual conjectures could be combined to produce a better forecast. 

Since Galton, combination forecasting was mathematically cast as an opinion pool.
Work in opinion pools began with Stone~\citep{stone1961opinion} in the early 1960s.
He assumed a set of experts had an agreed upon utility function related to decision making, and that experts could each generate a unique probability distribution to describe their perceived future "state of nature".
To build a single combined forecast, Stone proposed a convex combination of each expert's probability distribution over the future---an opinion pool.
Equally weighting individual predictions would reproduce Galton's model, and so the opinion pool was a more flexible way to combine expert opinions.

In the late 1960's, Granger and Bates formalized the concept of an optimal combination forecast.
In their seminal work~\citep{bates1969combination}, several methods were proposed for how to combine point predictions to reduce, as much as possible, the combined forecast's variance.
Methods for combinining forecasts was further advanced by Granger and Ramanathan, and framed as a regression problem~\citep{granger1984improved}.
Work by Granger, Bates, and later Ramanathan inspired several novel methods for combining point forecasts~\citep{gneiting2013combining,hora2015calibration,cooke1991experts,wallis2011combining}.
Combination forecasts often produce better predictions of the future than single models.

It wasn't until the 1990's that Cooke generalized the work of Stone and others, and developed an algorithm coined Cooke's method, or the Classical Model~\citep{cooke1988calibration,cooke1991experts} for combining expert judgment.
Every expert was asked to provide a probability distribution over a set of possible outcomes.
To assign weights to experts, a calibration score statistic compared the expert's probability distribution to an empirical distribution of observations.
Experts were assigned higher weights if their predictions closely matched the empirical distribution.
The calibration score was studied by Cooke and asymptotic properties were summarized based on Frequentist procedures~\citep{cooke1988calibration,cooke2015aggregation}.
Cooke's model also assigned experts a weight of $0$ for poor predictive performance, and
if an expert's performance was under some user-set threshold they were excluded from the opinion pool.
Cooke's model garnered much attention and has influenced numerous applications of combining expert opinion for forecasting~\citep{cooke2014validating,clemen2008comment,cooke2015aggregation}.

Alongside frequentist approaches to combination forecasting, Bayesian approaches began to gain popularity in the 1970's~\citep{morris1974decision}.
In the Bayesian paradigm, a decision maker (called a supra Bayesian), real or fictitious, is asked to evaluate expert forecasts and combine their information into a single probability distribution~\citep{hogarth1975cognitive,keeney1976group}.
The supra Bayesian starts with a prior over possible future observations and updates their state of knowledge with expert-generated predictive densities.
Combination formulas can be specified via a likelihood function $\ell$ meant to align expert-generated predictive densities with observed data.
The difficulties introduced by a Bayesian paradigm are familiar.
The choice of likelihood function and prior will affect how expert opinions are pooled. 
Past work proposed many different likelihood functions, and interested readers will find a plethora of examples in Genest and Zidek~\citep{genest1986combining}, and Clemen and Winkler~\cite{clemen1999combining,clemen1986combining,clemen1989combining}.

\subsection{Recent work in combination forecasting}

Recent work has shifted from combining point estimates to combining predictive densities.
Rigorous mathematical theory was developed and framed the problem of combining predictive densities~\citep{gneiting2013combining}.
Work combining predictive densities showed results similar in spirit to Granger and Bates'~\citep{bates1969combination,granger1984improved} work on combining point predictions.
Ranjan and Gneiting~\citep{ranjan2010combining,gneiting2013combining} showed a set of calibrated predictive distributions, when combined using a linear pool, necessarily leads to an overdispersed and therefore miscalibrated combined distribution.
This mimics Granger and Bates' results~\citep{bates1969combination}. 
They showed combining unbiased point predictions can lead to a combination method that makes biased point estimates.

This work in miscalibrated linear pools inspired new methods for recalibrating forecasts made from a combination of predictive densities. To recalibrate, authors recommend transforming the aggregated forecast distribution.
The Spread-adjusted Linear Pool~(SLP)~\citep{berrocal2007combining,glahn2009mos,kleiber2011locally} transforms each individual distribution before combining, the Beta Linear Pool~(BLP) applies a beta transform to the final combined  distribution~\cite{gneiting2013combining,ranjan2010combining}, and a more flexible infinite mixture version of the BLP~\cite{bassetti2018bayesian}, mixture of Normal densities~\cite{baran2018combining}, and empirical cumulative distribution function~\cite{garratt2019empirically} also aim to recalibrate forecasts made from a combination of predictive densities.

Machine learning approaches assume a broader definition of a model as any mapping that inputs a training set and outputs predictions.
This allows for more general approaches to combining forecasts called: ensemble learning, meta-learning, or hypothesis-boosting in machine learning literature.
Stacking and the super-learner approach are two active areas of machine learning research to combine models.
Stacked generalization~(stacking)~\citep{wolpert1992stacked} proposes a mapping from out-of-sample predictions made by models (called base-learners) to a single combination forecast.
The function that combines these models is called a generalizer and can take the form of any regression model, so long as it maps model predictions into a final ensemble prediction.
The super-learner ensemble takes a similar approach to stacking~\citep{van2007super,polley2010super}. 
Like stacking, the super-learner takes as input out-of-sample predictions from a set of models.
Different from stacking, the super-learner algorithm imposes a specific form for aggregating predictions, a convex combination of models, such that the weights assigned to each model minimize an arbitrary loss function that includes the super-learner predictions and true outcomes of interest.
By restricting how predictions are aggregated, super-learner is guaranteed better performance under certain conditions~\citep{van2007super,polley2010super}. 
Stacked and super-learner models often perform better than any individual forecasts and their success has led to applying them to many different problems~\citep{syarif2012application,sakkis2001stacking,che2011decision,wang2011comparative}, however the machine learning community is debating issues with stacked models~\cite{ting1999issues} and how they can be improved~\cite{dvzeroski2004combining}. 

\subsection{Open challenges in combination forecasting}
Combination forecasting has three distinct challenges: data collection, choice of combination method, and how to evaluate combination forecasts.

Crowdsourcing~\citep{howe2006rise,brabham2013crowdsourcing,abernethy2011collaborative,forlines2014crowdsourcing,moran2016epidemic} and expert elicitation~\cite{amara1971some,yousuf2007using,o2006uncertain} are two approaches to collecting judgemental forecasts that attempt to balance competing interests: the quantity and quality of judgemental predictions.
Crowdsourcing trades expertise for a large number of contributors.
Expert judgemental forecasting takes the opposite approach and focuses on a small number of independent high-quality forecasts. 
Both methods try to enlarge the space of potential predictions so that a combination method can create a more diverse set of predictive densities over future events~\citep{dietterich2002ensemble,bates1969combination}.

Combination methods are faced with developing a set of distributions over events of interest that take predictions as input and produce an aggregated prediction aimed at optimizing a loss function. 
Major challenges are how to account for missing predictions~\citep{capistran2009forecast}, correlated experts~\cite{armstrong1985long,bunn1985statistical,bunn1979synthesis}, and how to ensure the combination forecast remains calibrated~\cite{ranjan2010combining,gneiting2013combining,berrocal2007combining,glahn2009mos,kleiber2011locally,garratt2019empirically}.

No normative theory for how to combine expert opinions into a single consensus distribution has been established, and a lack of theory makes comparing the theoretical merits of one method versus another difficult. 
Instead, authors compare combination methods using metrics that measure predictive accuracy: calibration, and sharpness~\citep{jolliffe2012forecast,gneiting2007strictly,gneiting2011comparing,dawid2007geometry,hora2015calibration}.
Combination methods that output point forecasts are compared by measuring the distance between a forecasted point estimate and empirical observation.
Probabilistic outputs are expected to be calibrated and attempt to optimize sharpness, or the concentration of probability mass over the empirical observations~\citep{gneiting2007strictly,gneiting2011comparing,hora2015calibration,jolliffe2012forecast}.

\subsection{Past Reviews on Combination forecasting}

Our review underlines the digital age's impact on combination forecasting.
Collecting expert opinion in the past required one-on-one meetings with experts: in person, by phone, or mailed survey, and the internet decreased the burden of eliciting expert opinion by using online platforms to ask experts for their opinion~\citep{howe2006rise}.
Past work focused on using statistical models to combine forecasts, but increases in computing power broadened the focus from statistical models to machine-learning techniques.
Our review explores how the digital age transformed combination forecasting and is an updated look at methods used to aggregate expert forecasts. 

Many excellent past reviews of combination methods exist.
Genest and Zidek give a broad overview of the field and pay close attention to the axiomatic development of combination methods~\citep{genest1986combining}.
Clemen and Winkler wrote four reviews of aggregating judgmental forecasts~\citep{clemen1986combining, clemen1989combining, clemen1999combining,ISI:000081846900004}.
The most cited manuscript overviews behavioral and mathematical approaches to aggregating probability distributions, reviews major contributions from psychology and management science, and briefly reviews applications.
These comprehensive reviews center around the theoretical developments of combination forecasting and potential future directions of the science.
Our work is an updated, and more applied, look at methods for aggregating expert predictions.

\section{Methods}
\label{sec.methods}

\subsection{Search algorithm}

The Web of Science database was used to collect articles relevant to combining expert prediction.
The search string entered into Web of Science on 2019-03-06 was \textbf{(expert* or human* or crowd*) NEAR judgement AND (forecast* or predict*) AND (combin* or assimilat*)} and articles were restricted to the English language.
All articles from this search were entered into a database.
Information in this article database included: the author list, title of article, year published, publishing journal, keywords, and abstract (full database can be found at \url{https://github.com/tomcm39/AggregatingExpertElicitedDataForPrediction}).

To decide if an article was related to combining expert judgement, two randomly assigned reviewers (co-authors) read the abstract and were asked if the article was in or out of scope.
We defined an article as in-scope if it elicited expert judgments and combined them to make a prediction about natural phenomena or a  future event.
An article moved to the next stage if both reviewers agreed the article was in-scope.
If the two reviewers disagreed, the article was sent to a randomly assigned third reviewer to act as a tie breaker and was considered in scope if this third reviewer determined the article was in-scope.

Full texts were collected for all in-scope articles.
In-scope full texts were divided at random among all reviewers for a detailed reading.
Reviewers were asked to read the article and fill out a prespecified questionnaire~(Table~\ref{tab.prespecList}).
The questionnaire asked reviewers to summarize: the type of target for prediction, the methodology used, the experimental setup, and terminology associated with aggregating expert opinion.
If after a detailed review the article is determined to be out of scope it was excluded from analysis.
The final list of articles are called analysis-set articles.

\subsection{Analysis of full text articles}

From all analysis-set articles, abstract text was split into individual words, we removed English stop words---a set of common words that have little lexical meaning---that matched the Natural Language Toolkit (NLTK)'s stop word repository~\citep{loper2002nltk}, and the final set of non-stopwords were stemmed~\cite{willett2006porter}.

A univariate analysis: (i) counted the number of times a word $w$ appeared in abstract text per year~$n_{w}(t)$, (ii) the total number of words among all abstracts in that year~$(N_{t})$, and (iii) the frequency a word appeared over time $N_{w} = \sum_{t} n_{w}(t)$.
If a word $w$ did not appear in a given year it received a count of zero ($n_{w}(t)=0$).

Words were sorted by $N_{w}$ and a histogram was plotted of the top 5\% most frequently occurring words in abstract text.
Among the top $12$ most frequently occurring words, we plotted the proportion ($n_{w}(t)/N_{w})$ of each word over time.

Full text articles were scanned for key terms related to aggregating expert judgments.
Evaluation metrics, a preferred abbreviation, related names, whether the metric evaluated a binary or continuous target, and formula to compute the metric was included in a table~(Table~\ref{tbl.metrics}).  
Terms specific to aggregating judgmental data were grouped by meaning and listed in a table~(Table~\ref{tbl.terms}) along with a single definition.
If multiple terms mapped to the same concept, our preferred label was placed at the top. 

Frequencies and percents were computed for `Yes/No' prespecified questions related to analysis-set articles~(Statistics are presented in Table~\ref{tbl.question} and the list of all questions can be found in Table~\ref{tab.prespecList}).
Questions with text answers were summarized in the results. 

\section{Results}
\label{sec.results}

\subsection{Search results}

\begin{figure*}[ht!]
  \centering
  \includegraphics[scale=1.0]{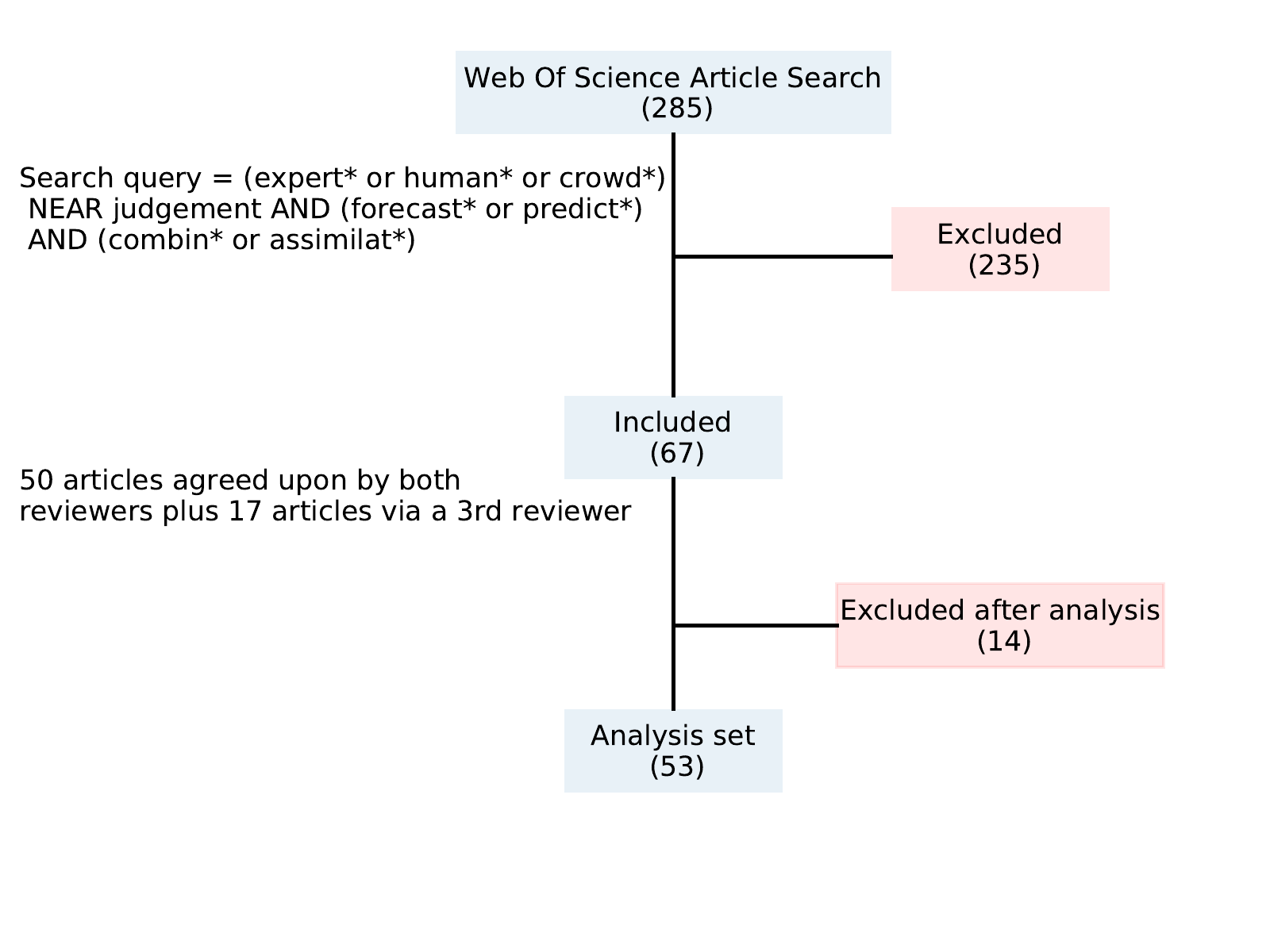}
  \caption{
A consort diagram for in-scope articles.
The search term used to collect the initial set of articles is reported and all intermediate steps between initial and analysis-set articles. 
  \label{fig.consort}}
\end{figure*}

The initial Web of Science search returned $285$ articles for review.
After random assignment to two reviewers, $218$ articles were agreed  to be out of scope. 
The most frequent reasons for exclusion were the lack of experts used for prediction or the use of experts to revise, rather than directly participate in generating, forecasts.
The $67$ in-scope articles come from $50$ articles two reviewers agreed to be in-scope, and $17$ out of $74$ articles a randomly assigned third reviewer considered in-scope.
Full text analysis determined another $14$ articles out of scope, and the final number of analysis-set articles was $53$~(Fig.~\ref{fig.consort}).

\begin{figure*}[ht!]
  \centering
  \includegraphics[scale=0.65]{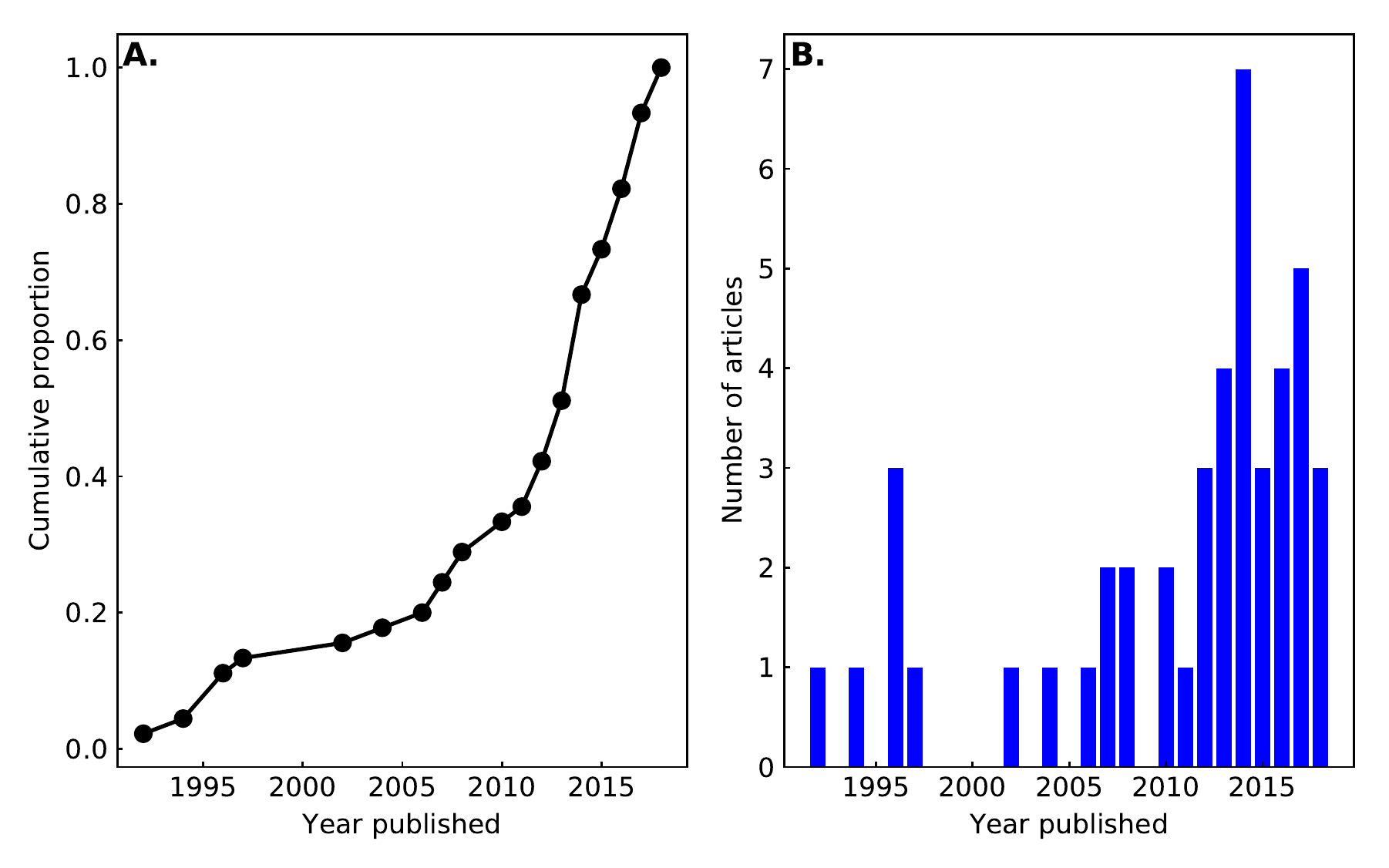}
  \caption{
  The cumulative proportion~(A.) and individual number~(B.) of articles published per year.
  The earliest in-scope article was published in $1992$ and most recent in $2018$.
  A sharp increase in publication occurred at or near $2010$.
  \label{fig.yearPublished}}
\end{figure*}

Analysis set articles were published from $1992$ to $2018$.
Publications steadily increase in frequency from $1992$ until $2011$.
After $2011$, publication rates rapidly increase until $2018$~(Fig.~\ref{fig.yearPublished}).

Analysis-set articles were published in $34$ journals, and the top publishing journals are: the \textit{International Journal of Forecasting}~($4$ articles), \textit{Reliability Engineering \& System Safety}~($3$ articles), and \textit{Risk Analysis} and \textit{Decision Analysis}~($2$ articles each).
Combination forecasting articles often emphasize the role of decision makers in forecasting, and
these top-publishing journals sit at the intersection of forecasting and decision sciences.

\begin{figure*}[ht!]
  \centering
  \includegraphics[scale=0.65]{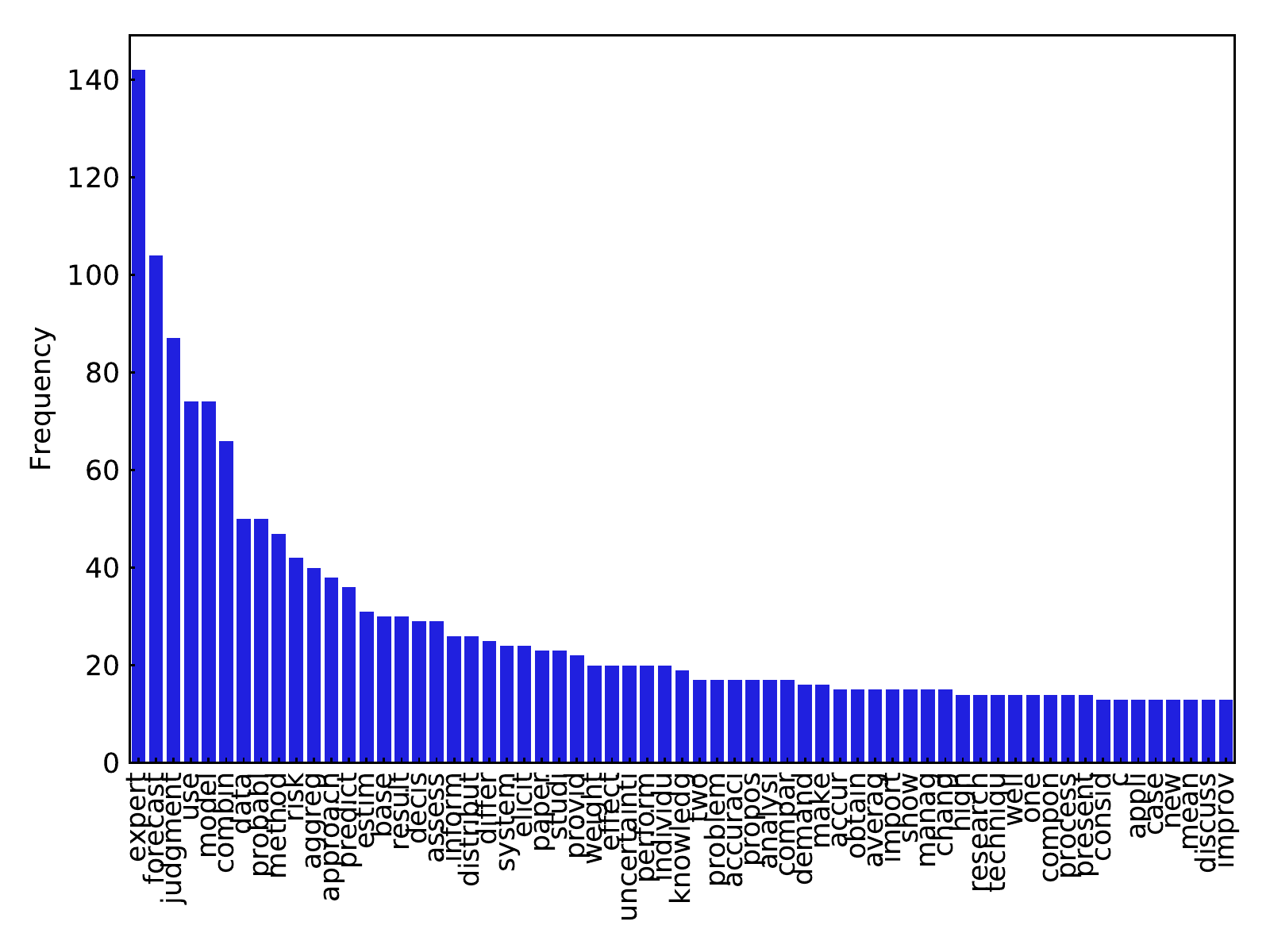}
  \caption{
The top 5 percent most frequent words used in all in-scope abstracts.
Expert, forecast, and judgment are the most frequent and likely related to the search words used to collect these articles.
  \label{fig.wordFreq}}
\end{figure*}

The top $10$ most frequent words found in articles' abstracts are related to our initial search: ``expert'',``judgment'', ``forecast'', ``combin'', and ``predict''.
Words related to modeling and methodology are also frequent: ``model'', ``method'', ``approach'', ``predict''.
The word ``assess'' appears less frequently in abstracts and the word ``accuracy'' even less frequent~(Fig.~\ref{fig.wordFreq}).

\begin{figure*}
    \centering
    \includegraphics[scale=0.75]{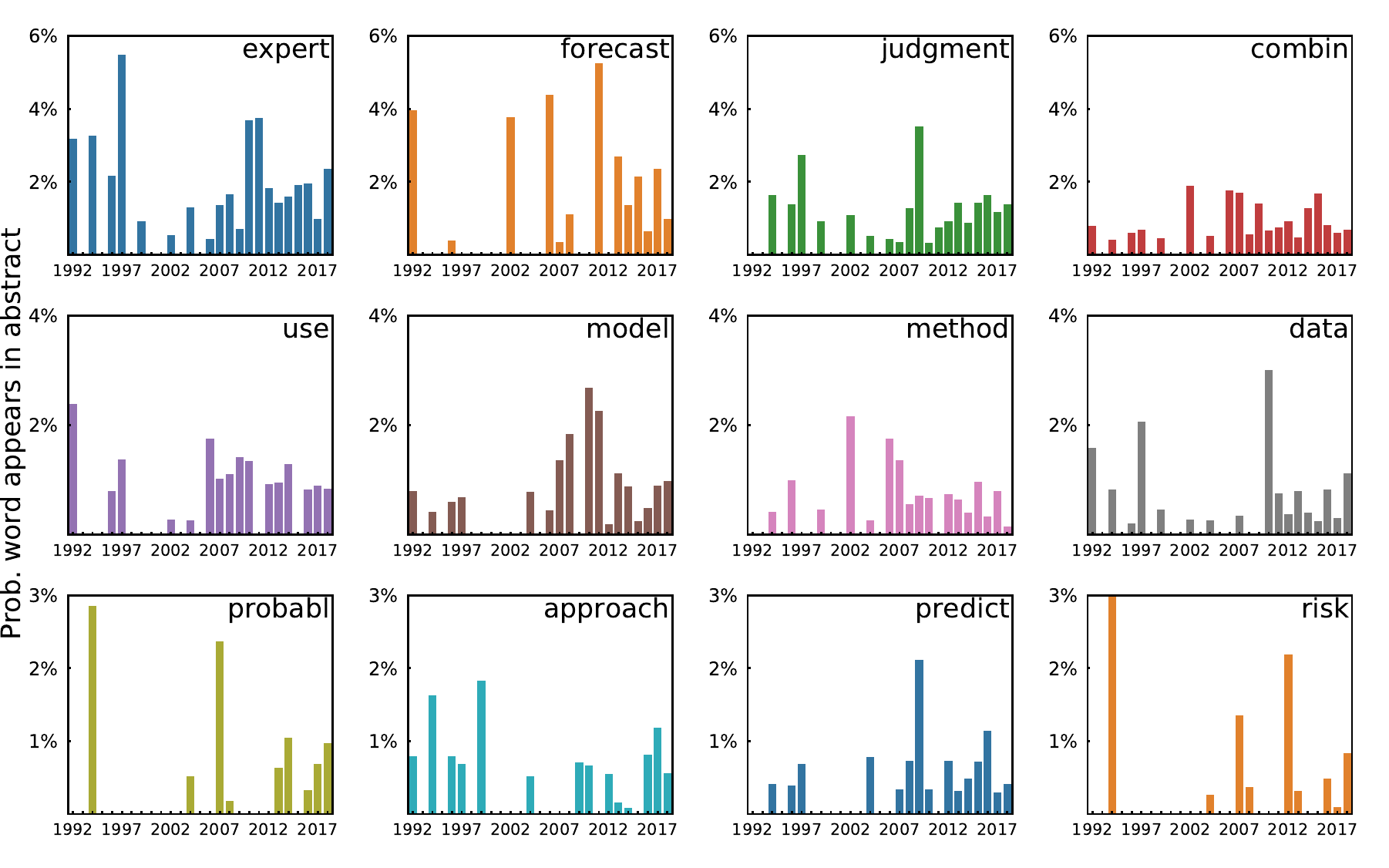}
    \caption{ 
    The annual proportion of the top $12$ most prevalent words among all abstract text. 
    For each year, word $w$ frequency was divided by the frequency of all words present in all abstracts.
    \label{fig.barplotCountsOverTime}}
\end{figure*}

The proportion of words: ``expert'', ``forecast'', ``model'', ``method'', and ``data'' appear intermittently in the $1990$s and appear more consistently in the $2000$s~(Fig.~\ref{fig.barplotCountsOverTime}).
The words ``probabili*'' and ``predict'' occur in abstract text almost exclusively after the year $2000$.
The rise of ``forecasts'', ``model'', and ``data'' suggests data-driven combination forecasting schemes may be on the rise, and the 
uptick of ``probabil*'' and ``predict'' could be caused by an increase in aggregating expert probability distributions (rather than point forecasts).

\subsection{Forecasting terminology}

Forecasting terminology centered around six distinct categories~(Table~\ref{tbl.terms}): frameworks for translating data and judgment into decisions~(Forecasting support system, probabilistic safety assessment), broad approaches to aggregating forecasts~(behavioral aggregation, mathematical combination, integrative judgment), specific ways experts can provide predictions~(integrative judgment, judgemental adjustment), terms related to weighting experts~(equal weighted linear pool, nominal weights), different names for classical models~(Cooke's method, mixed estimation), and philosophical jargon related to combination forecasting~(Laplacian principle of indifference, Brunswik lens model). 

Only a few concepts in the literature are assigned a single label, the majority are given multiple labels.
Some concepts' labels are similar enough that one label can be swapped for another.
For example, equal-weighted, 50-50, and unweighted all refer to assigning equal weights to expert predictive densities in a linear opinion pool. 
Other concepts are assigned different labels, for example forecasting support system and adaptive management, that may make it difficult to understand both terms refer to the same concept.

\subsection{Forecasting targets}

Forecasting research focused on predicting categorical variables~($18$ articles, 34\%) and time-series~($21$ articles, 40\%), but
the majority of articles attempted to predict a continuous target~($36$ articles, $68$\%)~(Table.~\ref{tbl.question}).

The type of forecasting target depended on the application.
Ecological and meteorological articles~\citep{johnson2018making,cooke2014out,li2012preliminary,tartakovsky2007probabilistic,morales2017characterization,borsuk2004predictive,abramson1996hailfinder,mantyka2014understanding,kurowicka2010probabilistic,wang2018bayesian} focused on continuous targets such as: the prevalence of animal and microbial populations, deforestation, and climate change.
Economics and managerial articles focused on targets like: the number of tourist arrivals, defects in programming code, and monthly demand of products~\citep{song2013combining,kabak2008aggregating,huang2016improving,failing2004using,shin2013robust}.
Political articles focused on predicting presidential outcomes, a categorical target~\citep{hurley2002combining,graefe2014accuracy,morgan2014use,graefe2015accuracy,graefe2018predicting,graefe2014combining}.
Risk-related targets were continuous and categorical: the probability of structural damage, nuclear fallout, occupational hazards, and balancing power load~\citep{klas2010support,zio1997accounting,cabello2012combination,adams2009acceptability,neves2008life,jana2019interval,hathout2016uncertainty,wang2008probabilistic,ren2002optimal,zio1996use,baecke2017investigating,brito2016bayesian,craig2001bayesian,mu1999multi,brito2012behavioral}.
Public health papers predicted continuous targets over time, like forecasting carcinogenic risk~\citep{evans1994use} and US mortality rates~\cite{alho1992estimating}.

Targets were often either too far in the future to assess, for example predicting precipitation changes in the next $1$ million years~\citep{zio1997accounting}, or related to a difficult-to-measure quantity, such as populations of animals with little or no monitoring~\cite{johnson2018making,borsuk2004predictive,mantyka2014understanding}.
The majority of analysis-set articles placed more importance on the act of building a consensus distribution than studying the accuracy of the combined forecast~\citep{johnson2018making,cooke2014out,li2012preliminary,klas2010support,zio1997accounting,song2013combining,clemen2007advances,tartakovsky2007probabilistic,morgan2014use,borsuk2004predictive,kabak2008aggregating,cabello2012combination,adams2009acceptability,neves2008life,failing2004using,evans1994use,hora2015calibration,abramson1996hailfinder,hathout2016uncertainty,wang2008probabilistic,mantyka2014understanding,kurowicka2010probabilistic,zio1996use,brito2016bayesian,gu2016expert,mu1999multi,wang2018bayesian,shin2013robust,brito2012behavioral,baron2014two}.

All articles defined a small number of specific forecasting targets.
The majority of targets related to safety.
Public health, ecology, and engineering applications focused on forecasting targets that, if left unchecked, could negatively impact human lives or the surrounding environment.
What differed between articles was whether the forecasting target could be assessed, and if ground truth data was collected on targets.

\subsection{Forecasting methodology}

Articles taking a Bayesian approach accounted for $25$\% of analysis-set articles and emphasized how priors can compliment sparse data~\citep{zio1997accounting,bolger2017deriving,ISI:000296286100010,tartakovsky2007probabilistic,huang2016improving,neves2008life,abramson1996hailfinder,ren2002optimal,mantyka2014understanding,brito2016bayesian,wang2018bayesian,brito2012behavioral}.
Many papers focused on assessing risk~\citep{zio1997accounting,brito2016bayesian,brito2012behavioral,tartakovsky2007probabilistic}.
For example, the risk of losing autonomous underwater vehicles was modeled using a Bayesian approach that incorporated objective environmental data and subjective probabilities of loss solicited from experts~\citep{brito2016bayesian,brito2012behavioral}.
Other papers assessed the impact of subsurface hydrology on water contamination~\citep{tartakovsky2007probabilistic}, the risk of structural deterioration~\cite{neves2008life}, and the economic risk associated with government expenditures~\cite{wang2018bayesian}.

Bayesian methods involved beta-binomial models, decision trees, mixture distributions, or Bayesian belief networks.
Often Bayesian models involved complicated posterior computations, requiring numerical integration to compute forecast probabilities.
Past work suggested a Bayesian framework could better model subjective probabilities elicited from experts~\citep{clemen2007advances}, however Frequentist techniques were used in almost 50\% of articles. 

Frequentist models for combining forecasts~\citep{cooke2014out,klas2010support,mak1996aggregating,hurley2002combining,morales2017characterization,borsuk2004predictive,hanea2018value,cabello2012combination,adams2009acceptability,alho1992estimating,evans1994use,jana2019interval,hora2015calibration,hathout2016uncertainty,wang2008probabilistic,ren2002optimal,kurowicka2010probabilistic,baldwin2015weighting,baecke2017investigating,seifert2013relative,gu2016expert,mu1999multi,graefe2014combining,alvarado2017expertise,shin2013robust,franses2011averaging} were typically convex combinations of expert judgment or linear regression models that included expert judgment as a covariate.
Including expert judgment as a covariate in a linear regression model is related to judgemental bootstrapping~\citep{armstrong2001judgmental} and the Brunswik lens model~\cite{hammond2001essential}. 
Both techniques are mentioned in analysis-set articles and rely on a Frequentist regression that divides human judgment into predictions inferred from data and expert intuition,
\begin{equation*}
  y_{e} | x_{e}, \beta_{0}, \beta, \sigma^{2}  \sim \mathcal{N}( \beta_{0} + \beta'x_{e} ,\sigma^{2})
\end{equation*}
where $y$ represents the expert's forecast, $\mathcal{N}$ is a Normal distribution, $x_{e}$ is a vector of objective information about the target of interest, $\beta$ are estimated parameters, and $\sigma^{2}$ is argued to contain expert intuition.
This model can then infer what covariates ($x_{e}$) are important to expert decision making and to what extent expert intuition ($\sigma^{2}$) is involved in prediction.

Articles that did not use classic regression combined statistical predictions (called `crisp') with qualitative estimates made by experts using fuzzy logic.
Cooke's method inspired articles to take a mixture model approach and weighted experts based on how well they performed on a set of ground-truth questions.

Articles using neither Bayesian or Frequentist models~\citep{johnson2018making,li2012preliminary,petrovic2006fuzzy,song2013combining,graefe2014accuracy,morgan2014use,cai2016simple,kabak2008aggregating,graefe2015accuracy,graefe2018predicting,failing2004using,ren2002optimal,hora2013median,baron2014two} resorted to: dynamical systems, simple averages of point estimates and quantiles from experts, and tree-based regression models.

The majority of models were parametric.
Non-parametric models included: averaging quantiles, equally weighting expert predictions, and weighting experts via decision trees.
These models allowed the parameter space to grow with increasing numbers of judgmental forecasts.
Parametric models included: linear regression, ARIMA, state space models, belief networks, the beta-binomial model, and neural networks.
Expert judgments, when combined and used to forecast, showed positive results in both nonparametric and parametric models.
Parametric Bayesian models and non-parametric models could better cope with sparse data than a parametric Frequentist model.
Bayesian models used a prior to lower model variance when data was sparse and non-parametric models could combine a expert judgments without relying on a specific form for the aggregated predictive distribution.

Authors more often proposed combining expert-generated point estimates compared to predictive distributions.
A diverse set of models were proposed to combine point estimates: regression models (linear regression, logistic regression, ARIMA, exponential smoothing), simple averaging, and neural networks~\citep{cabello2012combination,adams2009acceptability,mak1996aggregating,graefe2014combining,baron2014two}, and fuzzy logic~\cite{petrovic2006fuzzy,kabak2008aggregating,jana2019interval,ren2002optimal}.
Authors that combined predictive densities focused on simpler combination models.

Most predictive distributions were built by asking experts to provide a list of values corresponding to percentiles.
For example, a predictive density would be built by asking each expert to provide values corresponding to the 5\%, 50\% (median), and 95\% percentiles.
Combination methods either directly combined these percentiles by assigning weights to each expert density~\citep{ISI:000327676900001,hanea2018value,morales2017characterization,cai2016simple,ISI:000391078100005,kabak2008aggregating,zio1997accounting,BRITO201655}, or built a continuous predictive distribution that fit these discrete points~\cite{brito2012behavioral,abramson1996hailfinder,neves2008life,failing2004using,wang2008probabilistic,kurowicka2010probabilistic}.

\subsection{Forecasting evaluation metrics}
Only 42\% (22/53) of articles evaluated forecast performance using a formal metric.
Formal metrics used in analysis-set articles are summarized in Table~\ref{tbl.metrics}.
The articles that did not include a metric to compare forecast performance either did not compare combination forecasts to ground truth, evaluated forecasts by visual inspection, or measured success as the ability to combine expert-generated forecasts.
Among articles that did evaluate forecasts, most articles focused on point estimates~(68\%, 15/22) versus probabilistic forecasts~(23\%, 5/22), and two articles did not focus on point or probabilistic forecasts from experts.

The most commonly used metrics to evaluate point forecasts were: the Brier score, mean absolute (and percentage) error, and root mean square error.
Even when predictive densities were combined, the majority of articles output and evaluated point estimates.

A small number of articles combining probability distributions used metrics that evaluated aggregated forecasts based on density, not point forecasts.
Expert forecasts were evaluated using relative entropy and a related metric, the calibration score~(see Table~\ref{tbl.metrics} for details).
These metrics were first introduced by Cooke~\citep{cooke1988calibration,cooke1991experts}.

The logscore is one of the most cited metrics for assessing calibration and sharpness~\citep{gneiting2007strictly,gneiting2011comparing,hora2015calibration} for predictive densities, but was not used in any of the analysis-set articles.
Instead, analysis-set articles emphasized point estimates and used metrics to evaluate point forecasts.

Three articles conducted an experiment but did not use any formal metrics to compare the results.
Two articles used no evaluation and one article visually inspected forecasts.

\subsection{Experimental design}

Among all analysis-set articles, 22/53~(42\%) conducted a comparative experiment.
Most articles did not evaluate their forecasting methods because no ground truth data exists.
For example, articles would ask experts to give predictions for events hundreds of years in the future~\citep{zio1997accounting,zio1996use}.
Articles that didn't evaluate their combined forecast but did have ground truth data concluded that the predictive distribution they created was ``close'' to a true distribution.
Still other articles concluded their method successful if it could be implemented at all.

\subsection{Training data}
Rapidly changing training data---data that could change with time---appeared in 41\% of articles.
Data came from finance, business, economics, and management and predicted targets like: monthly demand of products, tourist behavior, and pharmaceutical sales~\citep{baecke2017investigating,petrovic2006fuzzy,wang2008probabilistic,klas2010support,franses2011averaging}.
In these articles, authors stress experts can add predictive power by introducing knowledge not used by statistical models, when the quality of data is suspect, and where decisions can have a major impact on outcomes. 
The rapidly changing environment in these articles is caused by consumer/human behavior.

Articles applied to politics stress that experts have poor accuracy when forecasting complex (and rapidly changing) systems unless they receive feedback about their forecast accuracy and have contextual information about the forecasting task~\citep{graefe2014accuracy,graefe2015accuracy,graefe2018predicting,satopaa2014probability}.
Political experts, it is argued, receive feedback by observing the outcome of elections and often have strong contextual knowledge about both candidates.  

Weather and climate systems were also considered datasets that rapidly change.
The Hailfinder system relied on expert knowledge to predict severe local storms in eastern Colorado~\citep{abramson1996hailfinder}.
Weather systems are rapidly changing environments, and this mathematical model of severe weather needed training examples of severe weather.
Rather than wait, the Hailfinder system trained using expert input.
Expert knowledge was important in saving time and money, and building a severe weather forecasting system that worked.

Ecology articles solicited expert opinion because of sparse training data, a lack of sufficient monitoring of wildlife populations, or to assign subjective risk to potential emerging biological threats~\citep{li2012preliminary,mantyka2014understanding,kurowicka2010probabilistic}

Manuscripts that explicitly mention the training data describe the typical statistical model's inability to handle changing or sparse data, and suggest expert predictions may increase accuracy~\citep{seifert2013relative,song2013combining}.

\subsection{Number of elicited experts and number of forecasts made}

\begin{figure*}[ht!]
  \centering
  \includegraphics[scale=0.65]{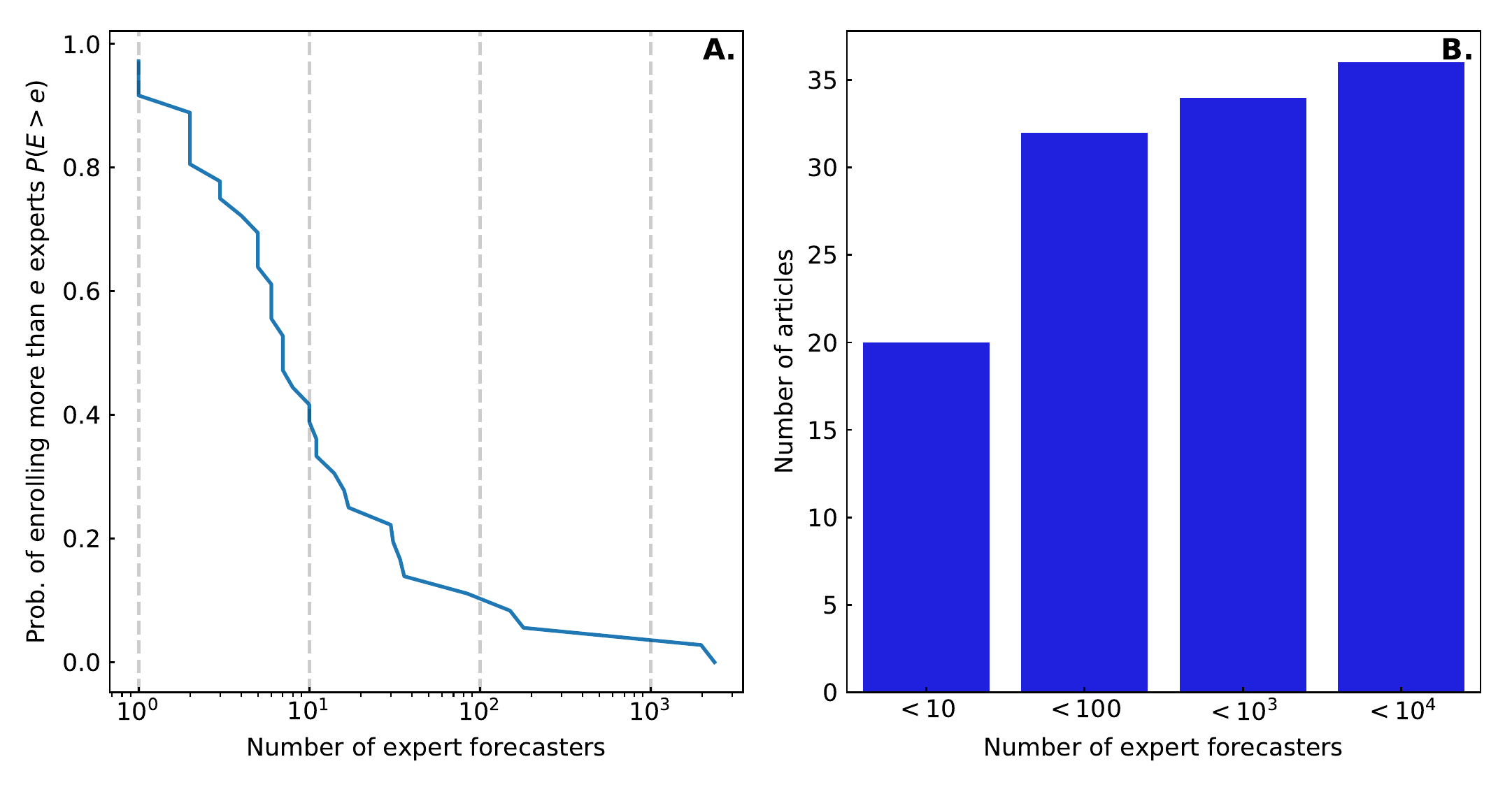}
  \caption{
The complimentary cumulative distribution (CCDF) of the number of experts elicited per article~(A.).
The proportion of articles enrolling less than 10, less than 100, less than 10$^{3}$, and less than 10$^{4}$ expert forecasters~(B.).
The small number of articles enrolling more than 10$^{3}$ were simulation studies.
  \label{fig.numForecasters}}
\end{figure*}

Over 50\% of articles combined forecasts from less than 10 experts.~(Fig.~\ref{fig.numForecasters}).
Several articles describe the meticulous book-keeping and prolonged time and effort it takes to collect expert judgments.
The costs needed to collect expert opinion may explain the small number of expert forecasters.

Two distinct expert elicitation projects produced articles that analyzed over $100$ forecasters.
The first project~\citep{seifert2013relative} asked experts from music record labels to predict the success (rank) of pop singles.
Record label experts were incentivized with a summary of their predictive accuracy, and 
an online platform collected predictions over a period of $12$ weeks.

One of the most successful expert opinion forecasting systems enrolled approximately $2000$ participants and was called the Good Judgement Project~(GJP)~\citep{mellers2014psychological,ungar2012good,satopaa2014probability}.
Over a period of $2$ years, an online platform was used to ask people political questions with a binary answer (typically yes or no) and to self-assess their level of expertise on the matter.
Participants were given feedback on their performance and how to improve with no additional incentives.
Both projects that collected a large number of forecasters have common features.
An online platform was used to facilitate data collection, and questions asked were simple, either binary (yes/no) questions or to rank pop singles. 
Both project incentivized participants with feedback of their forecasting performance.

\begin{figure*}[ht!]
  \centering
  \includegraphics[scale=0.65]{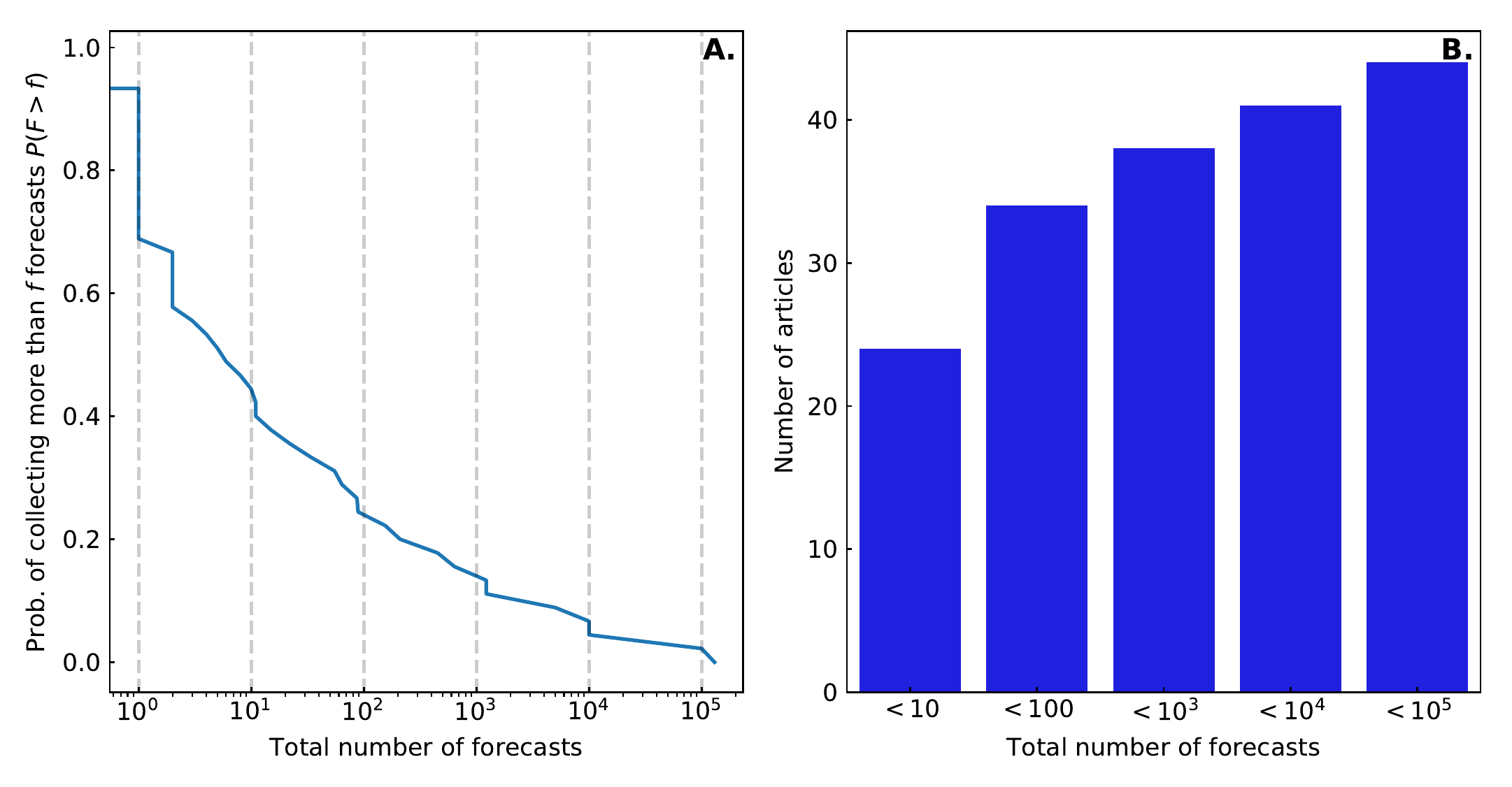}
  \caption{
Complimentary cumulative distribution of the total number of forecasts made per article~(A.), and 
the proportion of articles eliciting less than 10, 100, 10$^{3}$, 10$^{4}$, and 10$^{5}$ forecasts.
Articles collecting more than 10$^{4}$ forecasts were simulations.
  \label{fig.numForecasts}}
\end{figure*}

Close to 80\% of articles reported less than 100 total forecasts~(Fig.~\ref{fig.numForecasts}) and studies reporting more than $10^4$ forecasts were simulation based (except the GJP).
Recruiting a small number of experts did not always result in a small number of forecasts.
Authors assessing the performance of the Polly Vote system collected 452 forecasts from 17 experts~\citep{graefe2014accuracy,graefe2015accuracy,graefe2018predicting}, and
a project assessing the demand for products produced $638$ forecasts from $31$ forecasters~\citep{alvarado2017expertise}.

The time and energy required to collect expert opinion is reflected in the low number of forecasters. 
Some studies did succeed to produce many more forecasts than recruited forecasters, and they did so by using an online platform, asking simpler questions, and giving forecasters feedback about their forecast accuracy.

\section{Discussion}

\label{sec.discussion}

Combining expert predictions for forecasting continues to shows promise, however rigorous experiments that compare expert to non-expert and statistical forecasts are still needed to confirm the added value of expert judgement.
The most useful application in the literature appeals to a mixture of statistical models and expert prediction when data is sparse and evolving.
Despite the time and effort it takes to elicit expert-generated data, the wide range of applications and new methods show the field is growing.
Authors also recognize the need to include human intuition into models that inform decision makers.

In any combination forecast, built from expert or statistical predictions, there is no consensus on how to best combine individual forecasts or how to compare one forecast to another~(Table~\ref{tbl.metrics}).
In addition to methodological disagreements familiar to any combination algorithm, expert judgemental forecasts have the additional burden of collecting predictions made by experts.
The literature has not settled on how to define expertise and an entire field is devoted to understanding how experts differ from non-experts~\citep{dawid1995coherent,farrington2006nature,ericsson2007capturing,rikers2005recent,de2014thought}.
Methods for collecting data from experts that are unbiased and in the least time-consuming manner is also an area of open inquiry.
An investigator must spend time designing a strategy to collect data from experts, and experts themselves must make time to complete this prediction task. 
There is a vast literature on proper techniques for collecting expert-generated data~\citep{ayyub2001elicitation,yousuf2007using,powell2003delphi,normand1998eliciting,leal2007eliciting,martin2012eliciting}.
Expert elicitation adds an additional burden to combination forecasting not present when aggregating purely statistical models.

Combination forecasting literature reiterated a few key themes: (i) the use of human intuition to aid statistical forecasts when data is sparse and rapidly changing, (ii) including experts because of their role as decision makers, (iii) using simpler aggregation models to combine predictive densities and more complicated models to combine point predictions, and (iv) the lack of experimental design and comparative metrics in many manuscripts.

Many articles introduced expert judgment into their models because the data needed to train a statistical model was unavailable, sparse, or because past data was not a strong indicator of future behavior.
When training data was available, researchers typically used expert forecasts to supplement statistical models.
Authors argued that experts have a broader picture of the forecasting environment than is present in empirical data. 
If experts produced forecasts based on uncollected data, then combining their predictions with statistical models was a way of enlarging the training data. 
Expert-only models were used when data on the forecasting target was unavailable.
Authors argued context-specific information available to experts and routine feedback about their past forecasting accuracy meant expert-only models could make accurate forecasts.
Though we feel this may not be enough to assume expert-only models can make accurate forecasts, without any training data these attributes allow experts to make forecasts when statistical models cannot.

Applications varied, but each field stressed the reason for aggregating forecasts from experts was due to decision-making under uncertainty. 
For example: deciding on how a company can improve their marketing strategy, what choices and actions can affect wildlife populations and our environment, deciding on the structural integrity of buildings and nuclear power plants.
Numerous articles emphasized the role of decision making in these systems by naming the final aggregated forecast a decision maker.

A longer history of combining point forecasts~\citep{galton1907vox,bates1969combination,granger1984improved} has prompted advanced methods for building aggregated forecasts from point estimates. 
Simpler aggregation techniques, like linear pools, averaging quantiles, and rank statistics, were used when combining predictive densities.
Besides the shorter history, simple aggregation models for predictive densities show comparable, and often, better results than more complicated techniques~\citep{clemen1989combining,rantilla1999aggregation}.
The reasons why simple methods work so well for combining predictive densities is mostly empirical at this time~\citep{makridakis1983averages,clemen1989combining,rantilla1999aggregation}, but under certain scenarios, a simple average was shown to be optimal~\cite{wallsten1997evaluating,wallsten1997combining}.

A small percentage of research took time to setup an experiment that could rigorously compare combination forecasting models.
Most articles measured success on whether or not the combination scheme could produce a forecast and visually inspected the results.
In some cases visual inspection was used because ground truth data was not present, but in this case, a simulation study could offer insight into the forecasting performance of a novel combination method.
No manuscripts compared predictions between forecasts generated by experts only, a combination of experts and statistical models, and statistical models only.
Past research is still unclear on the added value experts provide statistical forecasts, and whether expert-only models provide accurate results.

To support research invested in aggregating expert predictions and improve their rigorous evaluation, we recommend the following: (i) future work spend more time on combining probabilistic densities and understanding the theoretical reasons simple aggregation techniques outperform more complicated models, and (ii) authors define an appropriate metric to measure forecast accuracy and develop rigorous experiments to compare novel combination algorithms to existing methods.
If not feasible we suggest a simulation study that enrolls a small, medium, and large number of experts to compare aggregation models.

Aggregating expert predictions can outperform statistical ensembles when data is sparse, or rapidly evolving.
By making predictions, experts can gain insight into how forecasts are made, the assumptions implicit in forecasts, and ultimately how to best use the information forecasts provide to make critical decision about the future.

\section{Funding}
This work has been supported by the National Institutes of General Medical Sciences (R35GM119582), the Defense Advanced Research Projects Agency, and the Centers for Disease Control and Prevention (1U01IP001122).
The content is solely the responsibility of the authors and does not necessarily represent the official views of NIGMS, the National Institutes of Health, the Defense Advanced Research Projects Agency, or and the Centers for Disease Control and Prevention.

\bibliographystyle{plainnat}
\section*{\refname}

\clearpage
\begin{longtable}{L{6cm} L{5cm} L{3cm}}
       \hline
        Related terms & Definition & Citations \\
        \hline
         Forecasting support system \hspace{35mm}
         Adaptive management          & 
         A framework for transforming data and forecasts into decisions.
         &
         \citep{alvarado2017expertise,song2013combining,baecke2017investigating,failing2004using,johnson2018making}
         \\
         \vspace{0.0625mm}
         
        (Probabilistic) Safety Assessment \hspace{35mm}
        (Probabilistic) Risk Assessment   & 
        A framework for investigating the safety of a system & 
        \citep{cooke2014out,zio1996use,zio1997accounting,jana2019interval,morales2017characterization,hanea2018value,hathout2016uncertainty,borsuk2004predictive,ISI:000296286100010,brito2012behavioral,kurowicka2010probabilistic,tartakovsky2007probabilistic,klas2010support,wang2018bayesian}\\
        \vspace{0.0625mm}
        
        Information set \hspace{35mm}
        Knowledge-base & Data available to an expert, group of experts, or statistical model used for forecasting. & 
        \citep{alvarado2017expertise,graefe2014accuracy,borsuk2004predictive,BRITO201655,abramson1996hailfinder,mak1996aggregating}\\
        \vspace{0.0625mm}\\
        
         Ill-structured tasks &  When changes to an environment impact the probabilistic links between cues an expert receives and their effect (how these cues should should be interpreted). & 
         \citep{seifert2013relative,huang2016improving}\\
        \vspace{0.0625mm}\\
         
         Behavioral aggregation\hspace{35mm}
         Behavioral combination\hspace{35mm}
         Structured elicitation & The support of expert discussion until they arrive at an agreed upon consensus distribution. & 
         \citep{hanea2018value,ISI:000296286100010,brito2012behavioral}\\
         \vspace{0.0625mm}\\    
         
         Mathematical combination \hspace{35mm}
         Mechanical integration      & The use of mathematical techniques to transform independent expert judgments into a single consensus distribution. & 
         \citep{ISI:000296286100010,petrovic2006fuzzy}\\
         \vspace{0.0625mm}\\    
         
         Judgmental adjustment
         Voluntary integration & Allowing experts to observe statistical forecasts, and provide their forecast as an adjustment to a present statistical forecast. & 
         \citep{alvarado2017expertise,huang2016improving,song2013combining,baecke2017investigating}\\
        \vspace{0.0625mm}\\    
         
         Integrative judgment
         Knowledge-aggregation & Forecasts from experts are incorporated into a forecasting model as a predictive variable. & 
         \citep{baecke2017investigating,mak1996aggregating}\\ 
         \vspace{0.0625mm}\\    
         
         Equal weighted  \hspace{35mm} 
         50-50 Weighting \hspace{35mm} 
         Unweighted & Assigning equal weights to all experts in a combination method. & 
         \citep{cooke2014out,hanea2018value,alvarado2017expertise,ISI:000327676900001,graefe2015accuracy}\\
         \vspace{0.0625mm}\\    
         
         Nominal weights & Weights obtained by assessing experts performance on a set of calibration questions, or on observed data.
         & \citep{baldwin2015weighting}\\
        \vspace{0.0625mm}\\     
         
         Cooke's method\hspace{35mm}
         Classical model & Combining expert opinion via a linear pool where weights depend on expert's answers to calibration questions with a known answer. &
         \citep{cooke2014out,morales2017characterization,zio1996use,hanea2018value,hathout2016uncertainty,ISI:000391078100005,ISI:000296286100010,brito2012behavioral,hora2015calibration, ISI:000327676900001}\\ 
        \vspace{0.0625mm}\\    
          
        Mixed estimation  \hspace{35mm}
        Theil-Goldeberger mixed estimation &  A method for combining expert and statistical forecasts, stacking statistical and expert point predictions into a single vector and fitting a linear regression model. & 
        \citep{alho1992estimating,shin2013robust}\\
        \vspace{0.0625mm}\\
        
        Laplacian principle of indifference \hspace{12mm}
        Principle of indifference & In the context of expert combination, having no evidence related to expert forecasting performance, models should weight experts equally. & 
        \citep{ISI:000391078100005}\\
        \vspace{0.0625mm}\\
        
        Brunswik lens model & 
        A framework for relating a set of criteria (or indicators), expert's judgment, and the "correct" judgment. & 
        \citep{seifert2013relative,franses2011averaging}\\
         \hline
             \caption{Terminology from analysis-set articles was collected and grouped by meaning.
              For each definition, the preferred terms is placed on top of all related terms.
              Definitions and preferred terminology were agreed upon by all coauthors.\label{tbl.terms} }\\
 
       \end{longtable}

\begin{table*}[ht!]
    \centering 
  \begin{tabular}{ L{5.5cm}cc }
    \hline
    Question                                                                    & Yes    & Total answers \\
                                                                                & N~(\%) & N \\
    \hline
    The primary target was\\
    \hspace{3mm}categorical                                                     &  18~(34)  & 53 \\ 
    \hspace{3mm}continuous                                                      &  36~(68)  & 53 \\
    \hspace{3mm}from a time series                                              &  21~(40)  & 52 \\
 \\
    A novel method/model was developed                                          & 26~(49)  & 53 \\   
    The authors implemented a\\
\hspace{3mm}Bayesian technique                                                  & 13~(25)  & 52 \\
    \hspace{3mm}Frequentist technique                                           & 26~(49)  & 53 \\ 
 \\   
    The model was\\
    \hspace{3mm}nonparametric                                                   & 13~(25) & 52 \\
    \hspace{3mm}parametric                                                      & 37~(73) & 51 \\
\\
    The model combined\\
    \hspace{3mm}point estimates                                                 & 29~(56) & 52 \\ 
    \hspace{3mm}probabilistic distributions                                     & 13~(37) & 52 \\
\\    
    Experts depended on data that could be updated, revised, or rapidly change  & 41 (21) & 51 \\
 \hline
  \end{tabular}
  \caption{
  A prespecified list of questions was asked when reviewing all in-scope articles.
  Frequencies and percentages were recorded for all binary questions.
  Questions a reviewer could not answer are defined as missing, causing some questions to have fewer than $53$ total answers.
  Answers to questions are on the article level and categories are not mutually exclusive.
  For example, an article could explore both a Frequentist and Bayesian model.
  \label{tbl.question}}
\end{table*}

\begin{sidewaystable}[ht!]
\begin{tabular}{L{4cm}C{3cm}C{4cm}cc}
    \hline
           &              &             & Binary or         &\\
    Metric & Abbreviation & Other names & Continuous target & Formula  \\
    \hline
    Absolute Loss  & AS                    & - & Categorical  &  $|P(F_{i}) - O_{i}|$\\
    Quadratic Loss & QS                    & - & Categorical  &  $\l[P(F_{i}) - O_{i}\r]^{2}$ \\
    Prediction Accuracy  & PA              & - & Categorical  &  $N^{-1} \sum_{i=1}^{N} \mathbbm{1}\l( F_{i} = O_{i} \r)$ \\
    Brier Score & BS                      & - & Categorical  &  $N^{-1} \sum_{i=1}^{N} \l[P(F_{i}) - O_{i}\r]^{2}$\\
    Mean Error & ME                       & - & Continuous   & $N^{-1} \sum_{i=1}^{N} (F_{i} - O_{i})$\\
    Mean Absolute Error & MAE             & Mean Absolute Deviation (MAD) & Continuous & $N^{-1} \sum_{i=1}^{N} |(O_{i} -F_{i} )|$\\
    Mean Absolute & MAPE    & Mean Percent Error (MPE)   & Continuous   & $N^{-1} \sum_{i=1}^{N} |(F_{i}/ O_{i} -1 )|$\\
    Percent Error                                       && Average percentage error (APE)\\
    Mean Squared Error & MSE              & - & Continuous & $N^{-1} \sum_{i=1}^{N} (F_{i} - O_{i})^{2}$ \\
    Root mean squared error & RMSE        & Root mean squared prediction error (RMSPE) & Continuous & $\sqrt{ N^{-1} \sum_{i=1}^{N} (F_{i} - O_{i})^{2}}$\\
    Proportion higher density & PHD       &  & Continuous & $N^{-1} \sum_{i=1}^{N} \mathbbm{1} \l\{ P\l[F(x_{i})\r] > P\l[G(x_{i})\r] \r\}$\\
    95\% Coverage probability & CP      & - & Continuous & $N^{1} \sum_{i=1}^{N} \mathbbm{1}\l( F_{2.5} < O_{i} < F_{97.5} \r)$ \\
    Judgemental Adjustment & JA        & - & Continuous & $\l(F_{i} - G_{i} \r)\Big/ G_{i}$ \\
    Forecast Improvement & FCIMP        & - & Continuous & $\l(|O_{i}-F_{i}| - |O_{i} - G_{i}| \r)\Big/ O_{i}$\\
    \hline
\end{tabular}
\caption{
\normalsize Metrics that in-scope articles used to evaluate both point and density forecasts.
A preferred term is listed (metric column), given an abbreviation and related names reported.
Whether the evaluative metric operates on a continuous or binary variable is stated and the computational formula presented. 
\label{tbl.metrics}}
\end{sidewaystable}

\begin{table}[ht!]
  \begin{tabular}{L{9cm}L{5cm}}
    \hline
    Question & Possible answers\\
    \hline
    \textbf{Forecasting target}\\
    Identify the primary predictive target?      & predictive target\\
    The primary target was categorical           & Y/N \\
    The primary target was continuous            & Y/N \\
    The primary target was from a time series    & Y/N \\
    Experts were given data related to the forecasting target? & Y/N \\

    \textbf{Terminology}\\
    List terms specific to aggregating crowdsourced data and quoted definition & term,def;term,def\\
    
    \textbf{Model}\\
    What models were used in forecasting?        & model$1$, model$2$, $\cdots$, model$n$\\
    Please list covariates included in any model & cov$1$, cov$2$, $\cdots$, cov$n$\\
    A novel model/method was developed                   & Y/N \\
    Did the authors implement a Bayesian technique?      & Y/N \\ 
    Did the authors use a Frequentist technique?         & Y/N\\
    Did the model account for correlation among experts? & Y/N \\
    The model combined point estimates                   & Y/N \\
    The model combined probabilistic distributions       & Y/N \\
    The model was parametric                             & Y/N \\
    The model was nonparametric                          & Y/N \\ 

    \textbf{Analysis data}\\
    Experts depended on data that could be updated, revised, or rapidly change? & Y/N \\
    
    \textbf{Experimental design}\\
    A comparative experiment was conducted               & Y/N\\
    How many expert forecasters were included?           & integer\\
    How many total forecasts were made?                  & integer\\
    What evaluation metrics were used?                   & metric$1$, metric$2$, $\cdots$, metric$n$\\

    \hline
  \end{tabular}
  \caption{ List of close-ended questions asked of each full-text article. Questions focus on the forecasting target, model, analysis data, and experimental design.\label{tab.prespecList}}
\end{table}

\graphicspath{{../_G/F1/}}

\graphicspath{{../_G/F2/}}

\graphicspath{{../_G/F4/}}

\graphicspath{{../_G/F3/}}

\graphicspath{{../_G/F5/}}

\graphicspath{{../_G/F6/}}

\end{document}